\documentstyle[11pt,epsf]{article}
    
\begin{document}

\large
\begin{center}
{\bf Quantum Hall Ferromagnetism in a Two-Dimensional Electron 
System\\}
\end{center}
\vspace*{0.1in}
\normalsize
\begin{center}
{J. Eom$^{1}$, H. Cho$^{1}$,  W. Kang$^{1}$, K.L. Campman$^{2}$,
A.C. Gossard$^{2}$, M. Bichler$^{3}$, and W. Wegscheider$^{3,4}$ \\}
\vspace*{0.1in}
{\it 1. James Franck Institute and Department of Physics,
  University of Chicago, Chicago, Illinois 60637, 2. Department of 
  Electrical Engineering, University of
California at Santa Barbara, Santa Barbara, California 93106, 
3. Walter Schottky Institut, Technische Universit\"{a}t
M\"{u}nchen, Am Coulombwall, D-85748 Garching, Germany,
4. University of Regensburg, Regensburg, Germany\\}
\end{center}

\vspace*{1in}

{\bf 
 Experiments on a nearly spin degenerate two-dimensional electron
  system reveals unusual hysteretic and relaxational transport in the
  fractional quantum Hall effect regime. The transition between the
  spin-polarized (with fill fraction   $\nu = 1/3$ ) and spin-unpolarized
  (  $\nu = 2/5$ ) states is accompanied by a complicated series of
  hysteresis loops reminiscent of a classical ferromagnet. In
  correlation with the hysteresis, magnetoresistance can either grow
  or decay logarithmically in time with remarkable persistence and
  does not saturate. In contrast to the established models of
  relaxation, the relaxation rate exhibits an anomalous divergence as temperature is reduced. These results
  indicate the presence of novel two-dimensional ferromagnetism with a complicated magnetic domain
  dynamic. }

The two-dimensional electron system (2DES) under low temperatures and high magnetic fields has become
a test bed for studying quantum phase transitions in low dimensions ({\it 1-3}). Unlike classical phase transitions
that are driven largely by thermal fluctuations, the phase transitions found in the quantum Hall effect (QHE)
regime represent a class of zero temperature phase transitions that are driven by strong electron-electron
interaction. Under intense magnetic fields, quantization of the electronic motion into Landau levels quenches
the kinetic energy and the interaction energy determines the thermodynamic properties of the underlying
2DES. At integral and certain fractional commensuration of the electron density and the applied magnetic
flux, called filling fraction , gain in the interaction energy produces transitions to the quantum Hall states
with Hall conductance $\sigma_{xy} = e^{2}/h$({\it 1, 2}). The collective, many-body nature of the QHE states is evident
from the dissipationless longitudinal transport and the existence of energy gap in the excitation spectrum.
The magnetotransport in the vicinity of a transition between two QHE phases is well described in terms of
critical behavior normally associated with a second-order phase transition. The universal scaling exhibited
by the QHE transitions and the existence of well-defined critical exponents provide compelling evidence
indicative of a zero-temperature quantum phase transition ({\it 3-5}). 

In addition to the transitions between two different QHE states, the translationally invariant quantum Hall
phases can exhibit novel forms of two-dimensional ferromagnetism ({\it 6}). Driven by gain in the anisotropy
energy over the Coulomb exchange energy, the ferromagnetic transitions in the multicomponent QHE
systems are accompanied by spatial ordering of pseudospin degrees of freedom represented by discrete
quantum numbers such as electronic spin, Landau level index, and electron layer quantum number.
Analogous to the ferromagnetically ordered electronic spins in magnetic systems, resulting quantum Hall
ferromagnets represent a new class of low-dimensional ferromagnets with remarkable properties ({\it 7, 8}). The
distinguishing features of quantum Hall ferromagnets include broken symmetry arising from spontaneous
magnetic ordering, existence of Goldstone mode in the ordered state, and topological objects as its
low-energy excitations ({\it 6}). 

The underlying symmetry of the quantum Hall ferromagnet is determined by the associated pseudospin of
the QHE phase. In case of even integral QHE states in single-layer electron systems, the spin configuration
of the associated Hall states becomes the pseudospin quantum number of the transformed state. Because the
possible spin geometries are limited to a fully polarized state with all electron spins aligned parallel to the
applied magnetic field and an unpolarized state with equal numbers of up and down spins, the resulting
pseudospin degree of freedom possesses a bimodal, Ising symmetry. In double-layer 2DESs found in double
quantum wells or a wide quantum well, the in-plane degree of freedom leads to a transition to a
two-dimensional XY ferromagnet ({\it 9-11}). In some QHE states, interplay of spin and the layer quantum
numbers can produce a transition to an Ising ferromagnet in wide-well system ({\it 8, 12}). 

Our experiment was performed on a high-quality GaAs/AlGaAs heterostructure. Data was taken at
pressures above 10 kbar where a large reduction in the magnitude of the electronic g factor favors
formation of spin-unpolarized fractional quantum Hall effect (FQHE) states. The evolution of the $\nu = 2/5$ 
FQHE state with pressure indicates enhancement of spin fluctuations against their tendency to align parallel
to the applied magnetic field. Under increasing pressure, the spin-unpolarized ground state competes against
the spin-polarized state, leading to coexistence over a broad region of pressure. The transition to the
pseudospin ferromagnetic state is distinguished by emergence of hysteretic transport and anomalous
temporal relaxation of magnetoresistance. The observed correlation between hysteretic and relaxational
evolution of magnetotransport points toward existence of intriguing domain dynamics in the pseudospin
ferromagnet.

The density of the GaAs/AlGaAs heterostructure used in the experiment was
$n$ = 3.5 $\times 10^{11}cm^{-2}$ with mobility of
$\mu = 2.4\times10^{6}cm^{2}/V sec$. A miniature beryllium-copper pressure cell was used to achieve high
pressure. Samples were immersed inside a homogeneous hydrostatic medium that transmits uniform
pressure to the sample. A small light-emitting diode was placed inside the pressure cell to illuminate the
sample at low temperatures. A gradual reduction in the electronic density was found with increasing
pressure. 

The magnetoresistivity of a high-quality 2DES sample is shown (Fig. 1) in the hysteretic region found
between 11 to 14 kbar of pressure. Although the FQHE states at $\nu$ = 2/3, 3/5, 1/2, and 1/3 are unchanged in
this region of pressure, the $\nu = 2/5$  FQHE state demonstrates an intriguing evolution with increasing
pressure. Between 11.2 and 13.5 kbar of pressure, the $\nu = 2/5$  FQHE state becomes progressively weaker at
higher pressures and exhibits strong hysteresis between up and down magnetic field sweeps. A slight
increase in the pressure to 13.8 kbar restores the $\nu = 2/5$  FQHE state. This type of reentrant behavior is also
found in other FQHE states and provides a compelling evidence of spin transitions that alter the spin
polarization of the associated FQHE states ({\it 13-17}). Tilted field experiments show that the electronic spins
are aligned completely parallel to the applied magnetic field for pressure below 13.5 kbar, whereas there
are equal numbers of up and down electronic spins in the state for pressure above 13.5 kbar ({\it 18}). 

In the temperature dependence of the normalized hysteretic resistance obtained around $\nu = 2/5$  for a 2DES
under 13 kbar of pressure (Fig. 2), the series of hysteretic resistance was obtained by subtracting the up and
down magnetoresistance sweeps and dividing it by the average magnetoresistance. The hysteresis profile was
found to be weakly dependent on the sweep rate. The hysteresis is most pronounced at the lowest
temperatures, and several crossings between up and down traces reveal a complicated hysteresis curve. The
fine features in the hysteresis resistance disappear at temperatures above 200 mK, and no hysteresis is
detected above 300 mK. The observed hysteretic behavior under pressure points toward presence of unusual
ground state in the vicinity of the $\nu = 2/5$  FQHE state. This is further reinforced by presence of logarithmic
time dependence of magnetoresistance.

An extended logarithmic time dependence is found over a range of fillings and temperatures for $\nu = 2/5$ 
(Fig. 3). The time scans were initiated after sweeping from zero field and holding the magnetic field at
  $\nu = 2/5$ . Typical time scans were performed over a 12- to 24-hour period, though scans lasting several days
have been performed on some occasions. The logarithmic time dependence is robust and reproducible under
repeated thermal and magnetic field cyclings. In all cases, the resistance at $\nu = 2/5$  increases logarithmically
in time and no saturation was ever detected. As the temperature is lowered, magnetoresistance relaxation
becomes enhanced and a larger change in the overall magnetoresistance was obtained. Similarly pronounced
relaxation behavior was found at different levels of excitation currents. To quantify the magnetoresistance
relaxation, we defined an effective relaxation rate in terms of the derivative 
of magnetoresistance ($R_{xx}$)
with respect to logarithm of time, d$R_{xx}$/d[log(t)]. The inset of Fig. 3 shows the effective relaxation rate
d$R_{xx}$/d[log(t)] as a function of temperature T at $\nu = 2/5$ . As the temperature is reduced, the relaxation
rate is found to increase dramatically. Least square analysis shows that 
the relaxation rate diverges as $1/T^{\alpha}$ with $\alpha$= 1.3.

The temporal evolution of magnetoresistance at fillings around $\nu = 2/5$  at a temperature of 69 mK (Fig. 4)
reveals a strong magnetic field dependence of the relaxation behavior. In particular, the logarithmic
relaxation of magnetoresistance is limited to fillings between $\nu $ = 3/7 and 1/3, and no time dependence is
found outside these filling factors. For fillings between $\nu $ = 3/7 and 2/5, magnetoresistance relaxes toward a
larger resistance, whereas for fillings between $\nu = 2/5$  and 1/3, magnetoresistance decreases as a function of
time. For both behaviors, no saturation in magnetoresistance was found.

Summarizing the magnetic field dependence of the relaxation rate d$R_{xx}$/d[log(t)] at 69 mK (Fig. 5A), a
sign change in the relaxation rate occurs at a magnetic field slightly above that corresponds to $\nu = 2/5$ . Most
remarkably, the magnetic field dependence of the relaxation rate reveals a striking similarity to the
hysteretic loop derived from magnetic field sweeps. Figure 5B illustrates the hysteresis in magnetoresistance
obtained by subtraction of up-sweep from the down-sweep data. The correlation between the hysteresis and
the logarithmic relaxation is highly suggestive of common origin for both hysteresis and the
magnetoresistance relaxation. In particular, the temporal evolution of magnetoresistance is suggestive of an
intriguing dynamics in the vicinity of the $\nu = 2/5$  FQHE state.

The theoretical studies of quantum Hall ferromagnetism have shown that certain quantum Hall states with
two nearly degenerate pseudospin Landau levels are susceptible to ferromagnetic ordering ({\it 7, 8}). In case of
the=2/5 FQHE state, the coexistence region of the two degenerate--polarized and unpolarized--spin
configurations under pressure can be mapped to an equivalent pseudospin ground state. Gain in the
anisotropy energy leads to spontaneous symmetry breaking and a transition to a two-dimensional easy axis
(Ising) ferromagnet ensues. Local potential fluctuations lead to nucleation of domains that are pinned by
defects, leading to complicated domain geometry. The anomalous magnetotransport observed in the vicinity
of the spin transition of the $\nu = 2/5$  FQHE state provides a compelling evidence of such a magnetic ordering
in the FQHE regime.

Near $\nu = 2/5$ , the magnetic field sweeps alter the energetics of the pseudospin interaction in the quantum
Hall ferromagnet, causing modification of the Zeeman splitting between competing spin configurations.
Consequently, it becomes necessary for individual pseudospin domains to grow, shrink, or coalesce in
response to the changes in the local energetics. Irreversible changes in the local spin configurations occur as
a result of complex domain dynamics. As the distribution of up and down pseudospin domains is altered,
history-dependent hysteresis accompanies the magnetic field sweeps, in close analogy with conventional
ferromagnets. 

The temperature dependence of hysteresis (Fig. 2) largely illustrates the strength of the magnetic ordering
at low temperatures. Due to the coupling between the electrical current and the pseudospin degree of
freedom through enhanced scattering at the domain boundaries, the hysteretic loops largely mirror the
temperature dependence of the order parameter in the quantum Hall ferromagnet. Coupled to the
pseudospin magnetization, the hysteresis gradually weakens and disappears at high temperatures. The ground
state below 300 mK may be best described as an ordered ferromagnet and at high temperatures, a
pseudospin paramagnet. Measurement of thermodynamic quantities such as specific heat ({\it 19, 20}) and
magnetization ({\it 21}) is expected to yield a similar behavior. 

The logarithmic time dependence exhibited by magnetoresistance in the vicinity of $\nu = 2/5$  is indicative of
relaxation of magnetic domains in the quantum Hall ferromagnet. The correlation between the relaxation
rate and the hysteresis likely arises from existence of energy-dependent interaction between the domains
within the sample. Depending on the nature of interaction and magnitude of the potential barrier, relaxation
between neighboring domains can occur at vastly different time scales ({\it 22, 23}). The motion between two
domains with identical magnetization involves crossing a region of opposite magnetization that acts as a
potential energy barrier. The domain wall energy of such a barrier is proportional to , where t
represents the anisotropy energy andis the exchange stiffness. Magnetic pressure exerted on the domain
walls produces slippage and propagation of domains through the sample. Such a relaxation process may be
generically described in terms of hopping between neighboring minima of potential wells that are driven via
thermal activation at finite temperatures ({\it 24}) and quantum tunneling near zero temperature ({\it 25-28}). 

In a large specimen possessing a distribution of various domain sizes and shapes, the collective response of
the magnetic domains can span an extended time range due to availability of response at all time scales. In
the vicinity of hard potentials in which two domains are separated by a large barrier, low probability of
tunneling and thermal activation produce slow relaxation. In contrast, the tendency for relaxation is
enhanced near soft potentials where domains can readily move and slide against each other. As a result,
presence of a broad distribution of domain energies manifests in a logarithmic time dependence instead of
exponential or power-law time dependence ({\it 22}). Although disorder can modify the relaxation behavior,
systems with strong residual interaction between neighboring domains exhibit slow, logarithmic relaxation
({\it23}).

Relaxation processes with time scale comparable to our system are found in spin glasses ({\it 29}) and
low-dimensional ferromagnets ({\it 30-32}). Aspects of relaxation such as randomness of local spin
configurations and broad distribution of relevant energy scales are common between these systems.
However, a salient feature of our experiment is the relaxation of magnetoresistance instead of magnetization.
This is accountable in terms of domain growth or shrinking ({\it 33}). As the scattering at domain walls enhances
the resistance, the fractional change in the overall magnetoresistance is proportional to the total length of the
domain wall in the system. The relaxation toward a larger resistance between= 3/7 and 2/5 (Fig. 4) may
be attributed to the overall growth of domain walls in the quantum Hall ferromagnet. On the other hand, the
relaxation toward a smaller resistance between $\nu = 2/5$  and 1/3 indicates overall reduction in the total length
of domain walls. The domain growth or shrinking in quantum Hall ferromagnets is presumably governed by
the Zeeman energy difference between neighboring pseudospin domains. Existence of domains in the sample
is also supported by strong nonlinearity in the current-voltage characteristics of the $\nu = 2/5$  FQHE ({\it 34}). 

One intriguing feature of the data is the temperature dependence of relaxation rate. The divergence of the
observed relaxation rate with temperature cannot be easily reconciled within a simple model of thermally
activated relaxation. In classical models of relaxations, thermally activated systems exhibit a relaxation rate
proportional to temperature ({\it 24}). In systems in which relaxations are driven by quantum tunneling, the
relaxation rate becomes independent of temperature at low temperatures ({\it 25-28}). Consequently, the
divergence in the temperature dependence of the relaxation rate points to some unconventional relaxation
process in the quantum Hall ferromagnet. Similar divergence in the relaxation rate at low temperature has
been observed in some low-dimensional ferromagnets ({\it 30-32}). Because there is no theoretical description of
relaxation in a quantum Hall systems at this time, it remains unclear what type of relaxation processes
dominate in a quantum Hall ferromagnet. Because disorder and Coulomb interaction play a nontrivial role
in the QHE, the possibility of some relaxation process unique to quantum Hall ferromagnets cannot be ruled
out at this time. 

In particular, coupling to nuclear spins may be responsible for the origin of the slow relaxation in the
vicinity of the $\nu = 2/5$  FQHE. In GaAs, the lattice nuclei consist of S = 3/2 nuclear isotopes of 69Ga and
71Ga and 75As with which conduction electrons interact through hyperfine coupling. In general, the
coupling between electronic and nuclear spins is extremely weak and the relaxations of nuclear spins occur
at a negligibly slow rate. However, presence of low-energy spin excitations has been shown to enhance the
coupling between the spins of electrons and nuclei in the quantum Hall regime. At= 1, presence of
electron spin texture excitation known as Skyrmions ({\it 1, 6}) leads to an enhanced nuclear spin diffusion and
gives rise to a large heat capacity at low temperatures ({\it 19, 20}). An earlier experiment also has shown that
nuclear spin polarization can produce hysteresis and memory effects in the magnetotransport ({\it 35}).
Consequently, enhanced coupling to nuclear spins cannot be ruled out in our experiment. A possible route of
coupling between nuclear and electronic spins is through existence of a Skyrmion-like excitation associated
with the $\nu = 1/3$  FQHE state. Because the $\nu = 2/5$  FQHE state can be considered to be a daughter state of the 
  $\nu = 1/3$  state, Skyrmionic excitations from the $\nu = 1/3$  FQHE state may be responsible for the coupling to the
nuclear spin in the $\nu = 2/5$ . Alternatively, the quasiparticles from the $\nu = 1/3$  FQHE that nucleate the $\nu = 2/5$ 
FQHE couple to the nuclear spins may be giving rise to the observed slow dynamics. Further experiments
will be necessary to elucidate the coupling to nuclear spin in the FQHE regime. 

In conclusion, we have observed transport and relaxational behavior associated with the ferromagnetic
ordering in the FQHE regime of a 2DES. Our experiment demonstrates a previously unknown type of
magnetic ordering of FQHE states and a slow, aging effect associated with multidomain structure. Further
experiments should reveal additional properties of this many-body magnetism. In particular, we speculate on
the nature of quantum Hall magnetism at micrometer or sub-micrometer lengthscales. Study of single
domain dynamics in the quantum Hall ferromagnets should reveal time-dependent phenomena such as
macroscopic quantum tunneling, switching, and telegraphic behaviors. Detection of such effects should lead
to greater understanding of many-body magnetism and quantum mechanics at small lengthscales.

\pagebreak 

REFERENCES AND NOTES

1. S. Das Sarma and A. Pinczuk, Eds., {\em Perspectives on Quantum Hall 
Effects} (Wiley, New York, 1997).\\
2. R. E. Prange and S. M. Girvin, Eds., {\em The Quantum Hall Effect,} (Springer-Verlag, New York, 1990).\\
3. S. L. Sondhi, S. M. Girvin, J. P. Carini, D. Shahar, {\it Rev. Mod. Phys.} {\bf 69}, 315 (1997).\\ 
4. H. P. Wei, D. C. Tsui, A. M. M. Pruisken, Phys.{\it Phys. Rev. B}
{\bf 33}, 1488 (1986).\\
5. H. P. Wei, D. C. Tsui, M. A. Paalanen, A. M. M. Pruisken, {\it Phys. Rev. Lett.} {\bf 61}, 1294 (1988).\\
6. S. M. Girvin and A. H. MacDonald, in ({\it 1}), pp. 161-224. 
7. T. Jungwirth, S. Shukla, L. Smrcka, M. Shayegan, A. H. MacDonald, {\it Phys. Rev. Lett.} {\bf 81}, 2328 (1998).\\
8. A. H. MacDonald, R. Rajaraman, T. Jungwirth, {\it Phys. Rev. B.} {\bf 60}, 8817 (1999).\\
9. S. Q. Murphy, J. P. Eisenstein, G. S. Boebinger, L. N. Pfeiffer, K. W. West, {\it Phys. Rev. Lett.} {\bf 72}, 728 (1994).\\
10. K. Yang {\it et. al., Phys. Rev. Lett.} {\bf 72}, 732 (1995).\\
11. A. Sawada {\it et. al., Phys. Rev. Lett.} {\bf 80}, 4534 (1998).\\
12. V. Piazza  {\it et. al., Nature} {\bf 402}, 638 (1999).\\
13. J. P. Eisenstein, H. L. Stormer, L. N. Pfeiffer, K. W. West, Phys. Rev. Lett. {\bf 62}, 1540 (1989).\\
14. R. G. Clark, et al., {\it et. al., Phys. Rev. Lett.} {\bf 62}, 1536 
(1989).\\
15. J. P. Eisenstein, H. L. Stormer, L. N. Pfeiffer, K. W. West, {\it Phys. Rev. B} {\bf 41}, 7910 (1990).\\
16. L. W. Engel, S. W. Hwang, T. Sajoto, D. C. Tsui, M. Shayegan, {\it Phys. Rev. B} {\bf 45}, 3418 (1992).\\
17. R. R. Du {\it et. al., Phys. Rev. Lett.} {\bf 75}, 3926 (1995).\\
18. H. Cho {\it et. al, Phys. Rev. Lett.}  {\bf 81}, 2522 (1998).\\
19. V. Bayot, E. Grivei, J. M. Beuken, S. Melinte, M. Shayegan, {\it Phys. Rev. Lett.} {\bf 79}, 1718 (1997).\\
20. S. Melinte, E. Grivei, V. Bayot, M. Shayegan, {\it Phys. Rev. Lett.} {\bf 82}, 2764 (1999).\\
21. S. A. J. Wiegers {\it et. al., Phys. Rev. Lett.} {\bf 79}, 3228 (1997).\\
22. J. J. Prejean and J. Souletie, {\it J. Phys. (Paris)} {\bf 41}, 1335 (1980).\\
23. D. K. Lottis, R. M. White, E. D. Dahlberg,  {\it Phys. Rev. Lett.} {\bf 67}, 362 (1991).\\
24. W. F. Brown, {\it Phys. Rev. } {\bf 130}, 1677 (1963).\\
25. E. M. Chudnovsky and L. Gunther,  {\it Phys. Rev. Lett.} {\bf 60}, 661 (1988).\\
26. M. Enz and R. Schilling, {\it J. Phys. C.} {\bf 19}, 1765 (1986).\\
27. J. L. van Hemmen and A. Suto, {\it Europhys. Lett.} {\bf 1} 481 (1986).\\
28. L. Gunther and B. Barbara, Eds., {\em Quantum Tunneling of the Magnetization} (North American
    Treaty Organization Advanced Study Institute, Kluwer, Dordrecht, Netherlands, 1995), series E,
    vol. 301. \\
29. E. Vincent, J. Hammann, M. Ocio, J. P. Bouchaud, L. F. Cugliandolo, 
in  {\it Complex behaviour of glassy systems: proceedings of the XIV Sitges conference}, 
Sitges, Spain, M. Rubi and    C. Perez-Vicente, Eds. (Springer-Verlag, Berlin, 1997), pp. 184-219 and 
references therein. \\
30. H. Yamazaki, G. Tatara, K. Katsumata, K. Ishibashi, Y. Aoyagi, {\it J. Magn. \&\ Magn. Mater.} {\bf 156}, 135 (1996).\\
31. J. E. Wegrowe, et al., {\it et. al., Phys. Rev. B} {\bf 52}, 3466 (1995).\\
32. J. E. Wegrowe, {\it et. al., Europhys. Lett.} {\bf 38}, 329 (1997).\\
33. F. Alberici-Kious, J. P. Bouchaud, L. F. Cugliandolo, P. Doussineau, A. Levelut, {\it Phys. Rev. Lett.} {\bf 81},
4987 (1998).\\
34. J. Eom et al., unpublished data. 
35. M. Dobers, K. v. Klitzing, J. Schneider, G. Weimann, K. Ploog, {\it Phys. Rev. Lett.} {\bf 61}, 1650 (1988).\\
36. We thank A. MacDonald and G. Mazenko for useful discussions. The work at the University of
    Chicago was supported in part by the Material Research Science and Engineering Center Program
    of the National Science Foundation (grant no. DMR- 9808595) and the David and Lucille Packard
    Foundation. 

\pagebreak

\begin{figure}
\epsfxsize=5.5in
\epsfbox{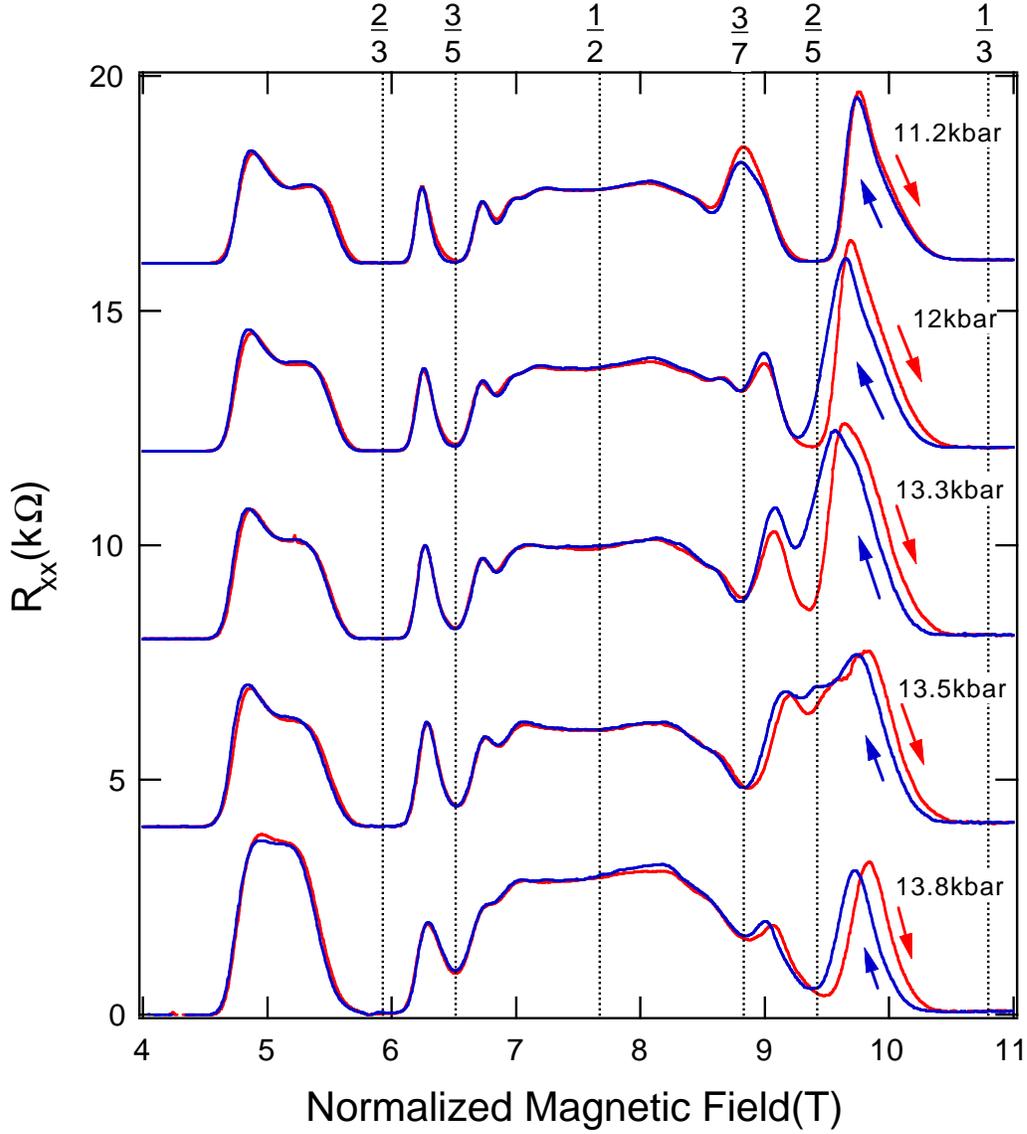}
\caption{Magnetoresistance of a high-quality GaAs/AlGaAs
heterostructure at 40 mK of temperature under pressure near filling
fraction  $\nu = 2/5$. Arrows indicate the sweep direction. Magnetic field
scale has been normalized to the highest pressure shown here for the
sake of comparison. Each pressure data point represents individual
cool-down after adjustments in the pressure.
\label{fig:rxxpresdep}}
\end{figure}

\begin{figure}
\epsfxsize=5.5in
\epsfbox{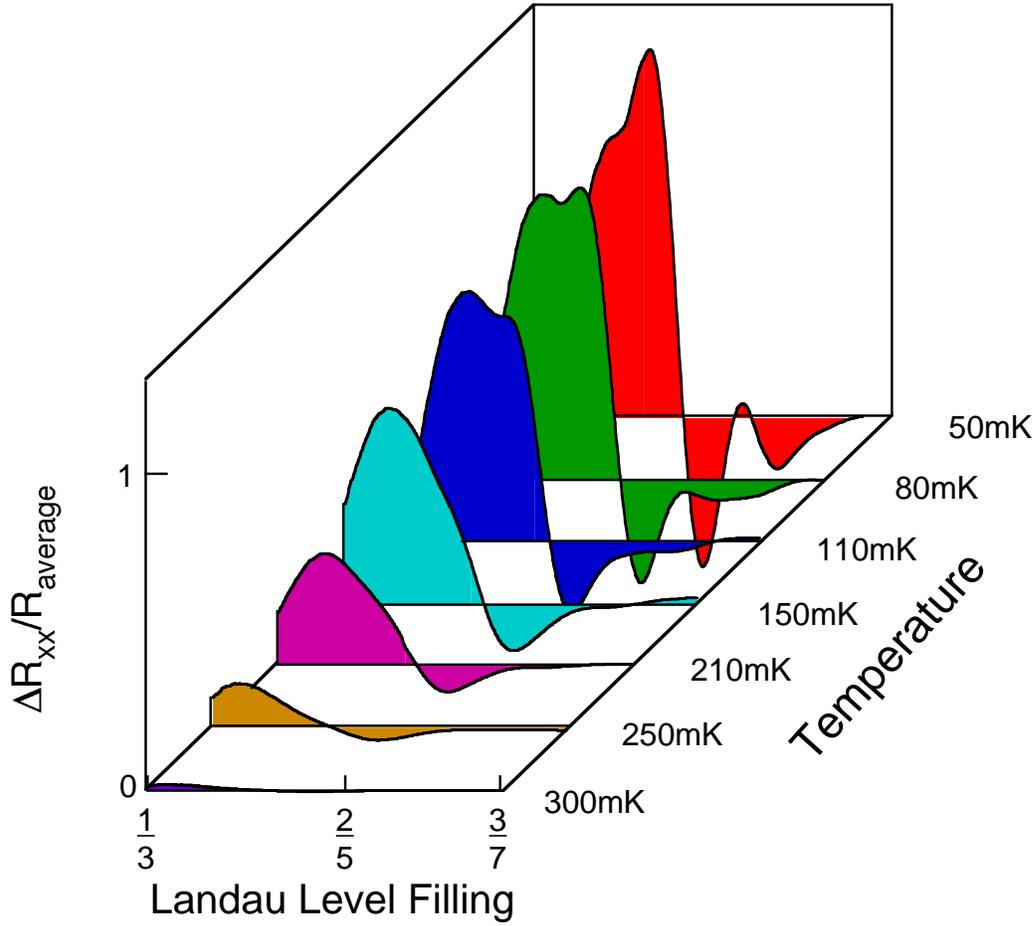}
\caption{Temperature dependence of normalized hysteresis resistance
around  $\nu = 2/5$  for a heterostructure sample under 13 kbar of
pressure. The hysteretic magnetoresistance was obtained by
subtracting the up and down magnetoresistance sweeps and dividing
it by the average magnetoresistance. 
\label{fig:hysteresis}}
\end{figure}

\begin{figure}
\epsfxsize=5.5in 
\epsfbox{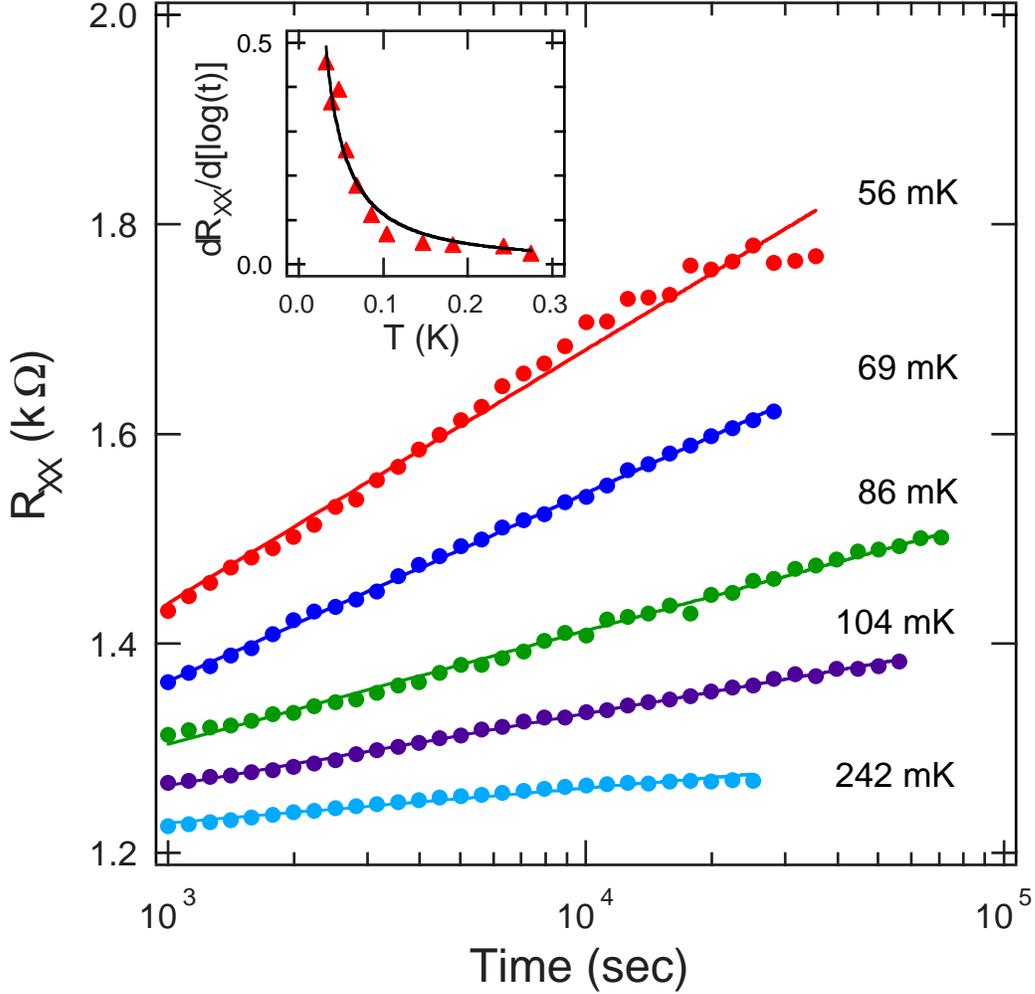}
\caption{Time evolution of the magnetoresistance at  $\nu = 2/5$  from a
GaAs/AlGaAs heterostructure sample under 13 kbar of pressure for
various temperatures. Time scans are initiated immediately after the
magnetic field sweep is finished. Because of slight eddy current
heating that occurs from magnetic field sweeps, a small temperature
drift (~2 to 3 mK) is detected during the initial ~20 min necessary to
equilibrate the temperature. Consequently, only the stable
temperature data after initial 1000 s is shown. Magnetoresistance has
been offset for the sake of clarity. (Inset) Plot of relaxation rate
d$R_{xx}$/d[log(t)]. The solid line represents a curve fit by 
$1/T^{\alpha}$, where $\alpha$ = 1.3.
\label{fig:mr-relax1}}
\end{figure}

\begin{figure}
\epsfxsize=5.5in
\epsfbox{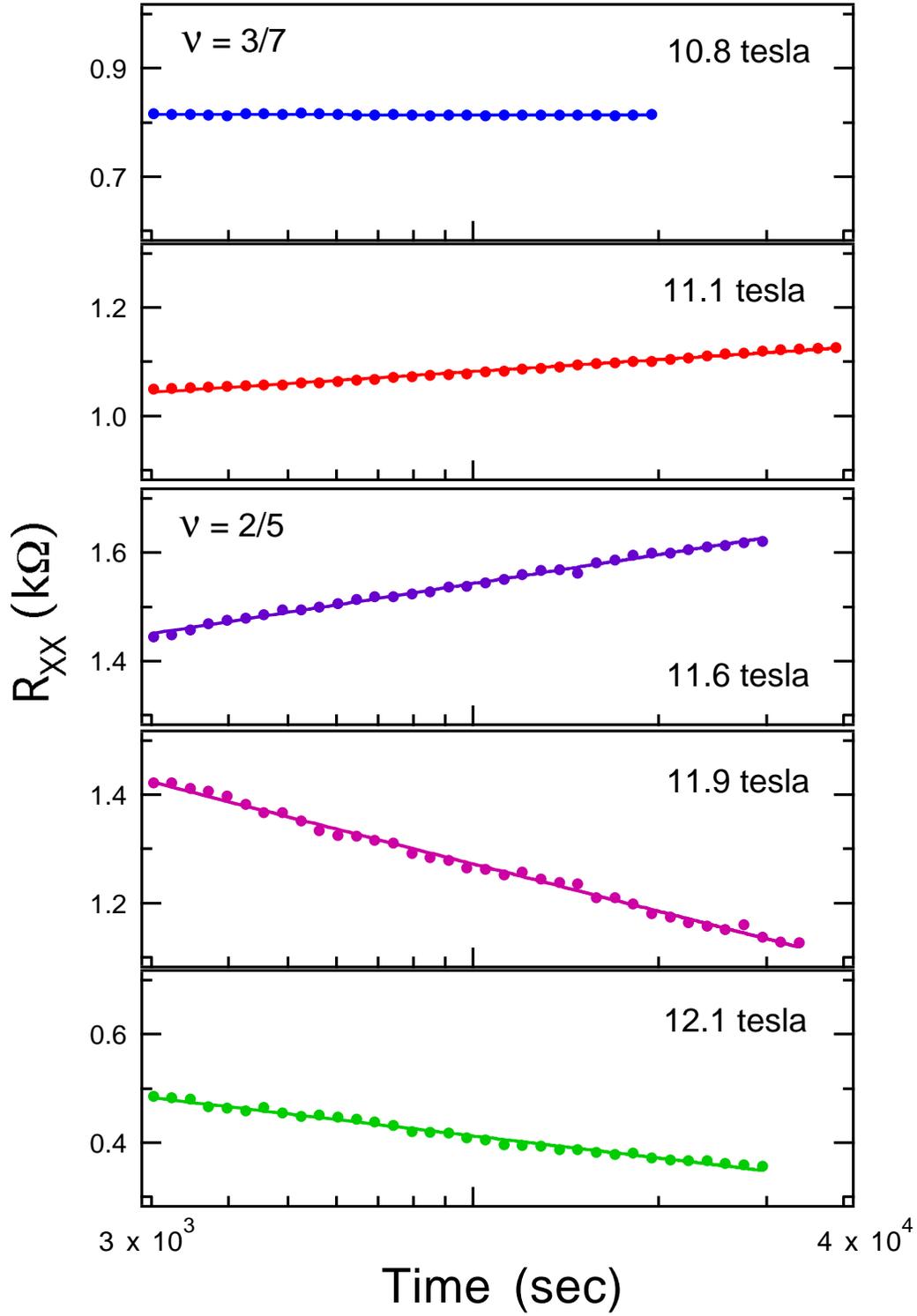}
\caption{Time evolution of magnetoresistance for different magnetic fields
around  $\nu = 2/5$ . For the sake of comparison, the same scale is used for different
magnetic fields. The time traces are taken at a temperature of 69 mK.
\label{fig:mr-relax2}}
\end{figure}

\begin{figure}
\epsfxsize=5.5in
\epsfbox{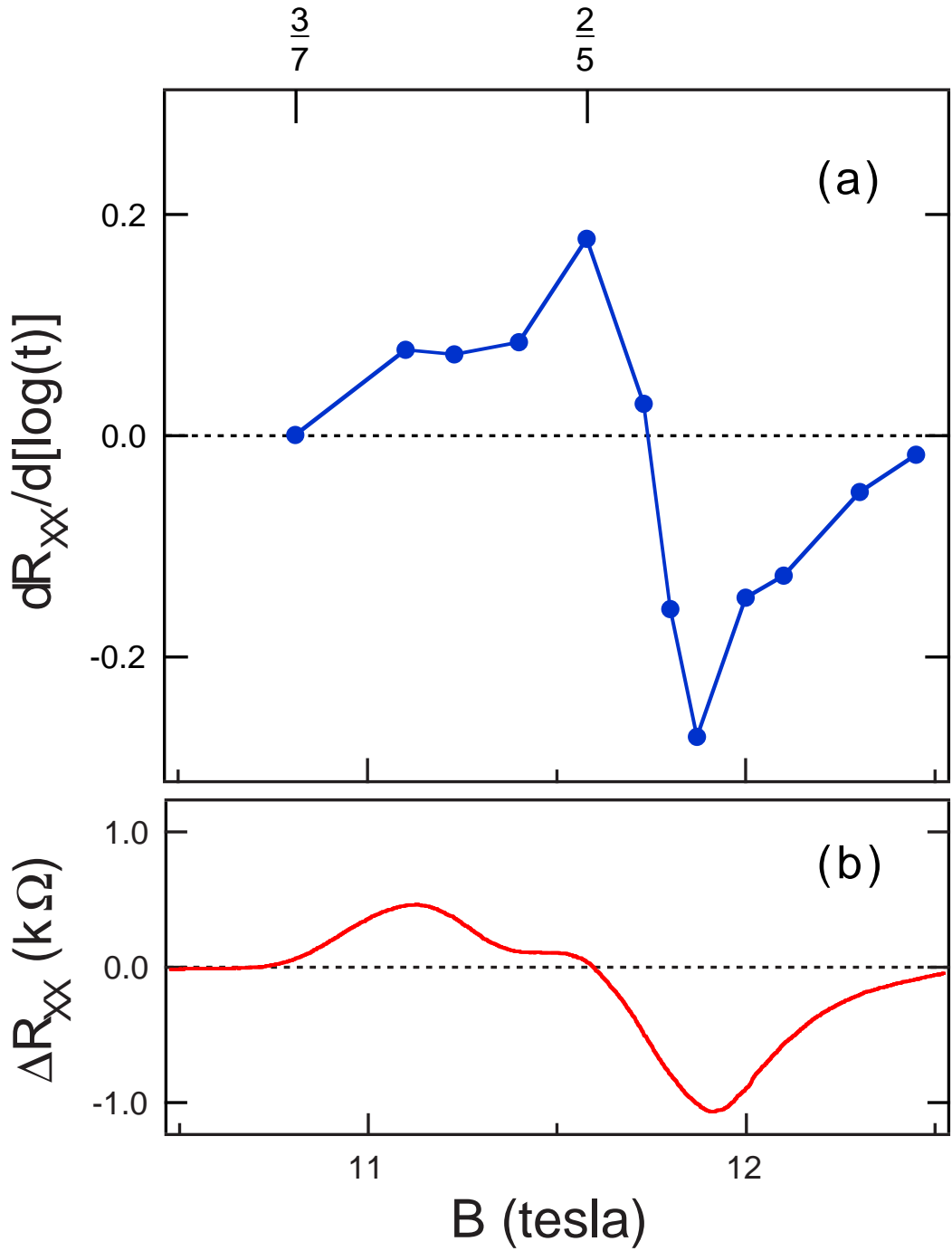}
\caption{(A) Magnetic field dependence of the relaxation rate at temperature
 of 69 mK. (B) Hysteresis in magnetoresistance, $R_{xx}$, obtained by
 subtracting the up- from down-sweep magnetoresistance around the $\nu = 2/5$ 
 fractional quantum Hall effect.
\label{fig:mr-relax3}}
\end{figure}

\end{document}